\documentclass{PoS}
\let\ifpdf\relax
\usepackage{bm}        % for math
\usepackage{amssymb}   % for math
\usepackage{amsmath}   % for math

\newcommand{\bea}{\begin{eqnarray}}
\newcommand{\eea}{\end{eqnarray}}
\newcommand{\beas}{\begin{eqnarray*}}
\newcommand{\eeas}{\end{eqnarray*}}

\newcommand{\ba}{\begin{align}}
\newcommand{\ea}{\end{align}}
\newcommand{\beq}{\begin{eqnarray}}
\newcommand{\eeq}{\end{eqnarray}}

\newcommand{\bi}{\begin{itemize}}
\newcommand{\ei}{\end{itemize}}

\title{Benchmarking the Bayesian reconstruction of the non-perturbative heavy $Q\bar{Q}$ potential}
\ShortTitle{Non-perturbative $Q\bar{Q}$ potential}
\author{\speaker{Y.~Burnier}\\%S 
        Institut de Th\'eorie des Ph\'enom\`enes Physiques, Ecole Polytechnique F\'ed\'erale de Lausanne, CH-1015, Lausanne, Switzerland,\\
        Albert Einstein Center for Fundamental Physics, Institute for Theoretical Physics, University of Bern, 3012 Bern, Switzerland\\
        E-mail: \email{yannis.burnier@epfl.ch}}

\author{A.~Rothkopf\\%S 
        Albert Einstein Center for Fundamental Physics, Institute for Theoretical Physics, University of Bern, 3012 Bern, Switzerland\\
        E-mail: \email{rothkopf@itp.unibe.ch}}

\abstract{
The extraction of the finite temperature heavy quark potential from lattice QCD relies on a
spectral analysis of the real-time Wilson loop. Through its position and shape, the lowest lying 
spectral peak encodes the real and imaginary part of this complex potential.
We benchmark this extraction strategy using leading order hard-thermal loop (HTL) calculations.
I.e. we analytically calculate the Wilson loop and determine the corresponding spectrum. By fitting
its lowest lying peak we obtain the real- and imaginary part and confirm that the knowledge of the
lowest peak alone is sufficient for obtaining the potential. We deploy a novel Bayesian approach to 
the reconstruction of spectral functions to HTL correlators in Euclidean 
time and observe how well the known spectral function and values for the real and imaginary part are reproduced.
Finally we apply the method to quenched lattice QCD data and perform an improved estimate of
both real and imaginary part of the non-perturbative heavy $Q\bar{Q}$ potential.
}

\FullConference{The 31st International Symposium on Lattice Field Theory - Lattice 2013,\\
		July 29-03, 2013\\
		Mainz, Germany}
		
\begin{document}
       
\section{Introduction}
In 1986, Matsui and Satz \cite{Matsui:1986dk} proposed the melting of $J/\Psi$ as signal for the deconfinement transition in heavy-ion collisions. The recent success of relativistic heavy-ion experiments \cite{Adare:2006ns,Tang:2011kr,Chatrchyan:2012np,Abelev:2012rv} in observing this phenomena for various charmonium and bottomonium states motivates us to develop a first principles description of heavy quarkonium melting.
      
Starting from quantum chromodynamics (QCD), the problem is simplified by the presence of a hierarchy of scales \cite{Brambilla:2004jw} $\Lambda_{QCD}\ll m_Q, \; T\ll {m_Q}$.
In this limit, the propagation of a heavy quark pair can be described by a rectangular temporal Wilson loop of time extent $t$ and spatial separation $r$ 
\begin{align}
 W_\square(t,r)=\frac{1}{N_c}\,{\cal P} {\rm Tr}[{\rm exp}[-ig\oint_\square dx^\mu A_\mu(x)]],
\end{align}
which satisfies a Schr\"odinger equation
\begin{align}
 i\partial_t W_\square(t,r) = \Phi(t,r) W_\square(t,r). \label{schro}
\end{align}
At late times, the function $\Phi(t,r)$ becomes time independent and defines the static potential:
\begin{align}
 V(r)=\lim_{t\to\infty} \Phi(r,t). \label{Eq:DefPot}
\end{align}
In QCD, the potential is difficult to calculate, even more so at finite temperature. In perturbation theory, the convergence of the expansion breaks down at large $r$. The hope that a high temperature would somewhat cure this issue did not materialize as only the leading order is fully known and higher orders are hard to calculate. In this paper we will discuss the possibility of using lattice simulations to compute the potential. There we face another problem, namely the fact that the lattice measurements are done in Euclidean time so that an analytic continuation to Minkowski space is needed.

In sec.~\ref{s2} we recall the important steps of the extraction of the $Q\bar Q$ potential from lattice measurements. In sec.~\ref{s3} we use leading order hard thermal loop (HTL) resummed perturbation theory to built a full test of the extraction. In sec.~\ref{s4} we test the procedure step by step and present in sec.~\ref{s5} the final results of the full procedure when applied to the HTL test data and to quenched lattice data. In sec.~\ref{s6} we conclude on the performance of the method and discuss the accuracy needed to be reached on the lattice.

%%%%%%%%%%%%%%%%%%%%%%%%
\section{Extracting the $Q\bar Q$ potential from lattice QCD}\label{s2}
%%%%%%%%%%%%%%%%%%%%%

To extract the potential $V(r)$ from Euclidean correlators we need to carry out the following steps \cite{Rothkopf:2011db,Burnier:2012az}:

\begin{enumerate}
 \item Calculate the Euclidean Wilson loop $W_\square(r,\tau)$ at several separation distances $r$ for all possible values along the imaginary time axis $\tau\in[0,\beta]$.
\item Connect the real-time Wilson loop to the measured Euclidean one, using the spectral representation:
\bea
 W_\square(r,t)=\int d\omega \; e^{-i\omega t}\; \rho_\square(r,\omega) \label{Eq:FourTrnasWL}\quad
\Leftrightarrow\quad
 W_\square(r,\tau)=\int d\omega \; e^{-\omega \tau}\; \rho_\square(r,\omega) \label{Eq:LapTransfWL}.
\eea
Calculate the most probable spectrum $\rho_\square(r,\omega)$ by inverting eq.\eqref{Eq:LapTransfWL}. This inversion can in principle be done with a bayesian inference method such as the MEM (see for instance \cite{Asakawa:2003re, Rothkopf:2011ef}) or our recent proposal \cite{Burnier:2013nla}.  
\item Extract the potential from $\rho(r,\omega)$. The first method would be by direct Fourier transform of the full $\rho_\square(r,\omega)$,
$$
 \quad V(r)={\displaystyle\lim_{t\to\infty}} \frac{ \displaystyle \int d\omega \;\omega \; e^{-i\omega t} \; \rho(r,\omega)}{\displaystyle \int d\omega \; e^{-i\omega t} \; \rho(r,\omega)}\label{Eq:DefPotSpec}.
$$
This requires a very good knowledge of the complete spectrum and is numerically very difficult.     
A better method was derived in ref.~\cite{Burnier:2012az}, were we showed in a fully general setup that the spectral function behaves as a skewed Lorentzian at small frequencies:
\begin{eqnarray}
\rho_\square(r,\omega)\notag&=&\frac{1}{\pi}e^{{\rm Im}[\sigma_\infty](r)} \frac{|{\rm Im}[V](r)|{\rm cos}[{\rm Re}[\sigma_\infty](r)]-({\rm Re}[V](r)-\omega){\rm sin}[{\rm Re}[\sigma_\infty](r)]}{ {\rm Im}[V](r)^2+ ({\rm Re}[V](r)-\omega)^2}\\&&+c_0(r)+c_1(r)({\rm Re}[V](r)-\omega)%+c_2(r)({\rm Re}[V](r)-\omega)^2
+\cdots\label{Eq:FitShapeFull}
\end{eqnarray}
hence it is sufficient to fit the lowest lying peak varying the parameters (${\rm Re}V,~{\rm Im}V,~ {\rm Re}(\sigma_\infty)$, ${\rm Im}(\sigma_\infty),~c_0,~c_1,~\dots$) of this functional form.
\end{enumerate}
Before applying blindly the method to lattice data, we will test all these steps in a simple case where all the corresponding quantities can be calculated explicitly.

%%%%%%%%%%%%%%%%%%%%%
\section{Potential and spectrum form HTLs}\label{secHTL}\label{s3}
%%%%%%%%%%%%%%%%%%%%%%

To benchmark the procedure, we use HTL resummed perturbation theory, where we can calculate  at leading order all quantities encountered in the steps 1-3 of the previous section. Note that on the lattice it is often easier to measure the Wilson line correlator (in a fixed gauge, for instance Coulomb) instead of the Wilson loop, thus we will study this quantity as well:
\begin{enumerate}
\item The Wilson loop contains a cusp divergence. Instead of renormalizing it, we introduce a cutoff $\Lambda$ to mimic the lattice regularization. After doing this it can be calculated numerically as well as the Wilson line correlator, which is also finite in the limit $\Lambda\to\infty$  \cite{Burnier:2013fca} (fig.~1).
\begin{figure}[h]
\centering
\includegraphics[width=7.5cm]{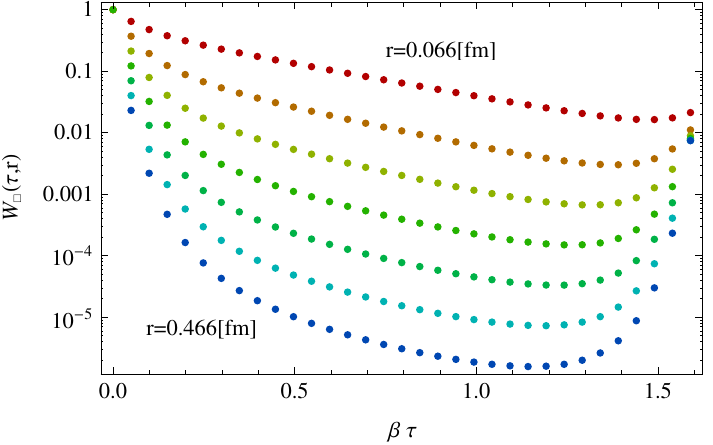}
\includegraphics[width=7.5cm]{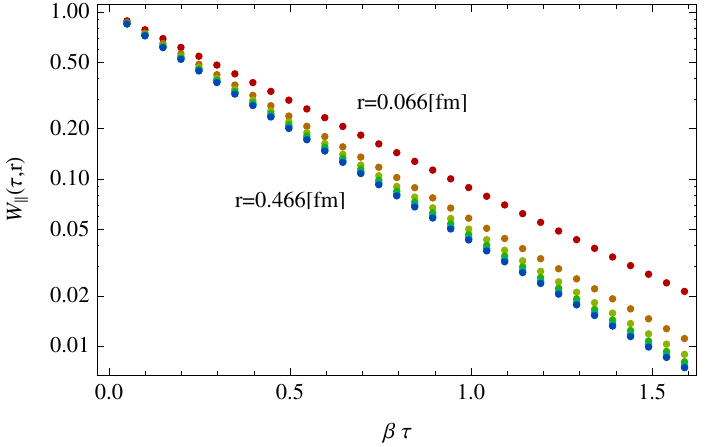}
\caption{(left) The Euclidean HTL Wilson loop $W^{\rm HTL}_\square(\tau,r)$ with momentum regularization $\Lambda=5\pi$ GeV evaluated at $T=2.33\times 270$ MeV in steps of $\Delta r=0.066$fm. (right) Wilson lines in the same case. }
\label{1}
\end{figure}

\item The spectral function can be calculated \cite{Burnier:2013fca} analytically at low frequency. In this domain, it has the same shape for Wilson loop and lines:
\bea
\rho^\Lambda_{\square}(r,\omega)&\propto&\rho^\Lambda_{||}(r,\omega)\propto
\frac1\pi \frac{|{\rm Im}[ V](r)|\cos\delta-({\rm Re}[ V](r)-\omega)\sin\delta}{({\rm Im}[ V](r))^2+({\rm Re}[ V](r)-\omega)^2}
\eea
with $\delta=\frac{|{\rm Im}[ V](r)|}{2T}$ the skewing of the Lorentzian. The full spectrum requires to numerically Fourier transform the real time correlator; results are shown in fig.~2.
\begin{figure}[h]
\includegraphics[width=7.5cm,angle=0]{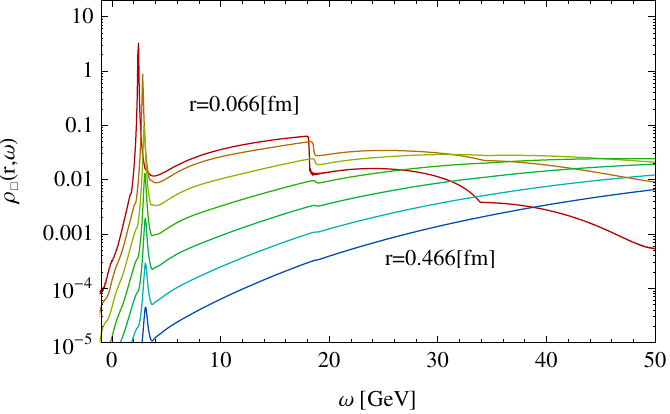}
\includegraphics[width=7.5cm,angle=0]{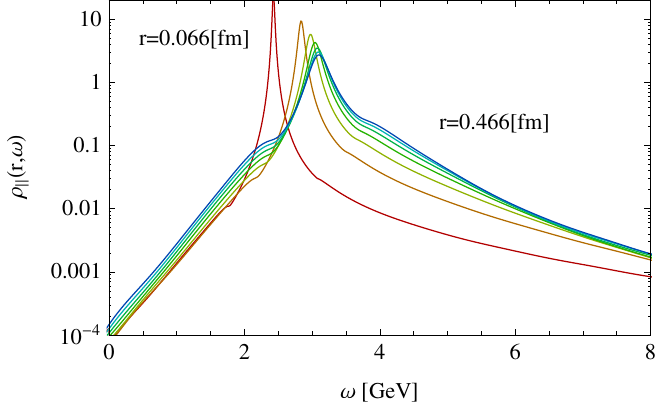} 
\caption{The spectral function of the HTL Wilson Loop (left) and Wilson lines correlator in Coulomb gauge (right). Note the sharp peaks and the large backgrounds.}   \label{2}
\end{figure}
\item The potential is well known \cite{Laine:2006ns,Beraudo:2007ky,Brambilla:2008cx} and is the same for Wilson loop and lines:
\bea
 &&V^{HTL}_{\square}(r)= V^{HTL}_{||}(r)=-\frac{g^2 C_F}{4\pi} \Big[ m_D + \frac{e^{-m_Dr}}{r} - i T\phi(m_D r)\Big]\label{VHTL}\\
&&\textrm{with   }\quad \phi(x)=2\int_0^\infty dz \frac{z}{(z^2+1)^2}\Big[1-\frac{\sin[zx]}{zx}\Big].\notag
\eea

\end{enumerate}

%%%%%%%%%%%%%%%%%%%%%%%%
\section{Test of the procedure}\label{s4}
\begin{figure}[h]
\centering
\includegraphics[width=6cm]{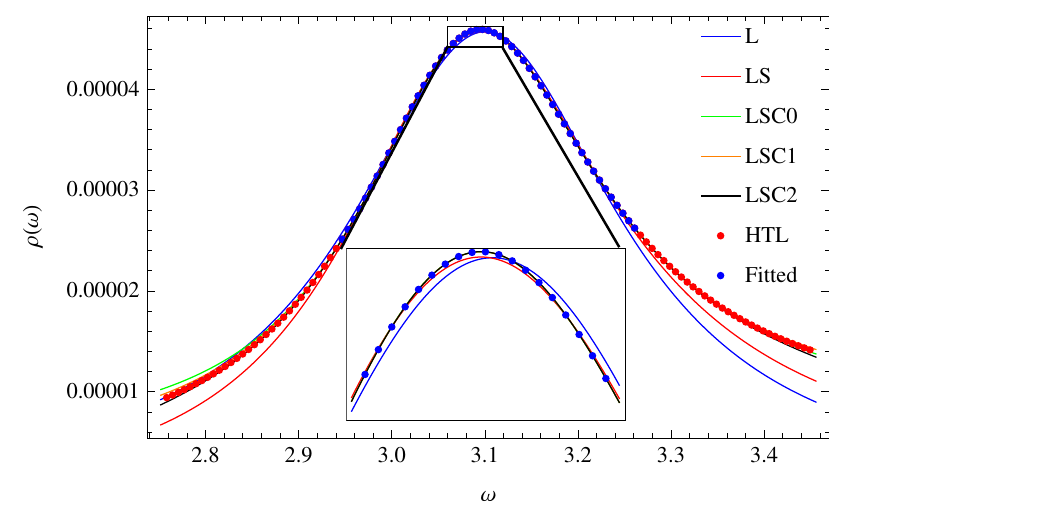}\hspace{-1cm}
\includegraphics[width=4.7cm]{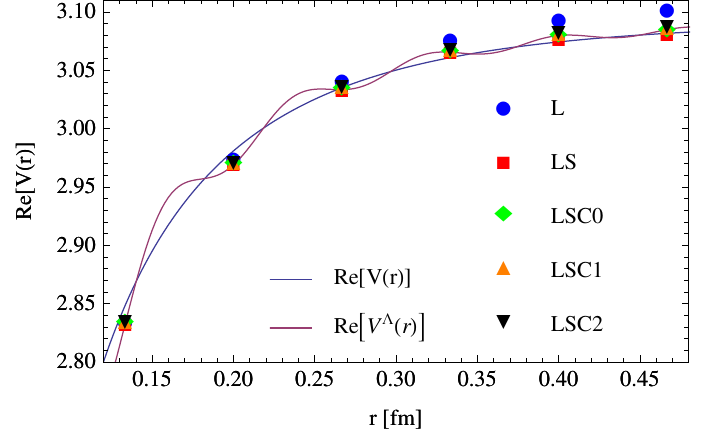}\hspace{0cm}
\includegraphics[width=4.6cm]{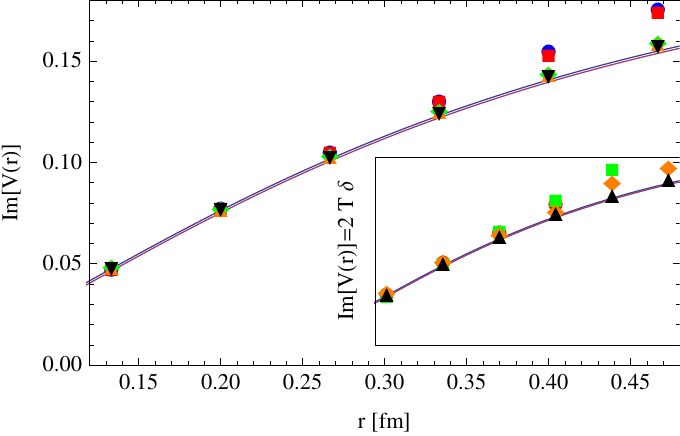}
 \caption{(left) Fits to HTL spectral function $\rho^\Lambda_\square(r,\omega)$ at  $r=0.49{\rm {\rm fm}}$ (right) with a naive Lorentzian (L), a skewed Lorentzian (LS) and a skewed Lorentzian with additional polynomial  terms (LSC0, LSC1, LSC2). Only the blue points (labeled ''Fitted'') are used for the fits and hence needed to compute the potential. (middle) Real and (right) imaginary part of the UV regularized ($\Lambda=5\pi {\rm GeV}$) HTL heavy quark potential (red, solid line) at $T=2.33T_C$ as well as the potential without cutoff (gray, dashed line). The various symbols denote the extracted values from fits based on a Lorentzian (L), skewed Lorentzian (LS) and a skewed Lorentzian with background terms (LSC0, LSC1, LSC2).} \label{Fig:SpecFits}
\end{figure}

%%%%%%%%%%%%%%%%%%%%%%%%%%
Having now the HTL spectra (fig.~2.) we can use the fitting procedure (point 3. of sec \ref{s2}) to get the potential.
%%%%%%%%%%%%%%%%%%%%%%%%%
%\subsection{Extracting the potential by Fitting the lowest peak}
%We fit the peaks of the HTL spectral function  shown in fig.~\ref{2} using the functional form (\ref{Eq:FitShapeFull}). 
Fitting the spectrum of the Wilson loop for each values of $r$, we get the potential (see fig.~\ref{Fig:SpecFits}).
The HTL potential being known, the results of the fit can be checked:
The simple Lorentzian fit (L) used in ref.~\cite{Rothkopf:2011db} leads to an overestimation of the potential at large $r$. This partly explains the counterintuitive results obtained there. The skewed fits reproduce the potential correctly. The incertitude coming from the fitting procedure are negligible and can be estimated by varying the number of parameters included in the fit and the fitting range.

%%%%%%%%%%%%%%%%%%%%%%%%%%
%\subsection{Test of the Analytical continuation}

We now turn to the analytical continuation, i.e. actually reconstructing the spectral function from the Euclidean correlator data. Obtaining the spectral function from the Euclidean correlator is a difficult problem. Here the model independent method of refs.~\cite{Cuniberti:2001hm, Burnier:2012ts, Burnier:2011jq} is not applicable as the Wilson loop is not periodic. The usual way is to resort to the standard or extended MEM \cite{Asakawa:2003re,Rothkopf:2011ef}. This method is tested in the following but has several issues. It cannot resolve the width of the lowest lying peak \cite{Burnier:2013fca} so that the imaginary part of the potential is far off and the real part not so precise. The peak also does not have a Lorentzian shape so that the fit is not stable and more worrisome, it shows only marginal improvements with better data and is numerically too expensive with very good data.

We therefore went again through all the steps of the derivation of the MEM and constructed a new Bayesian method solving many problems of the usual MEM \cite{Burnier:2013nla}. The main changes are: an unrestricted search space, where each point in the frequency domain is varied independently; a new entropy functional that does not contain asymptotically flat directions (unlike the Shanon-Jaynes entropy used in MEM) and finally, the hyperparameters are integrated explicitly (no Gaussian approximation). 

From the HTL Euclidean correlator (see fig.~1) calculated for an arbitrary number of points (32 to 256) to the tenth digit, we reconstruct the spectral function using the extended MEM where we have chosen to give the exact points (best case) and our new method where we added relative gaussian errors ($10^{-2}$ to $10^{-5}$) to the Euclidean data.

\begin{figure}[h]
\centering
\includegraphics[width=50mm,angle=-90]{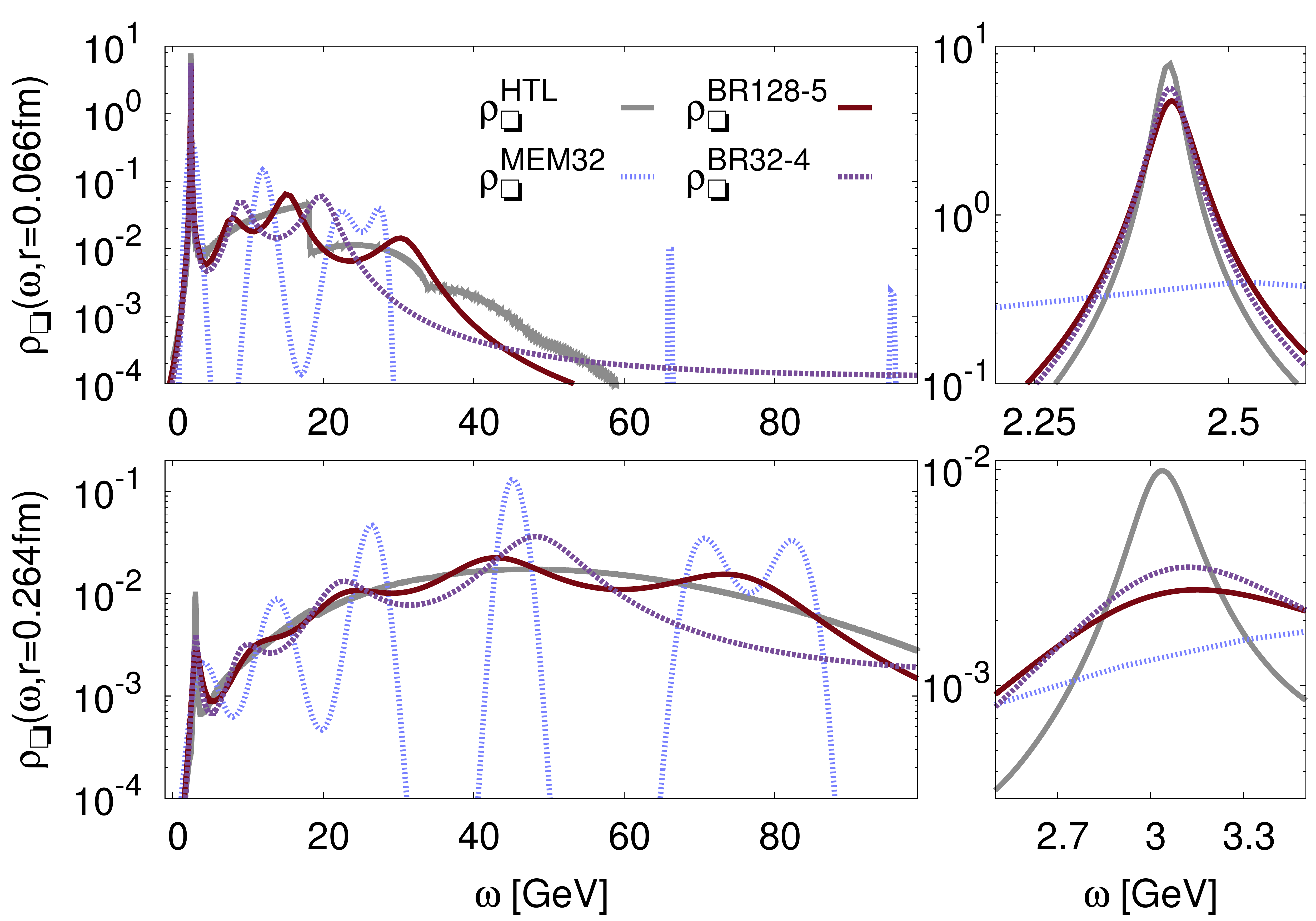}
\includegraphics[width=50mm,angle=-90]{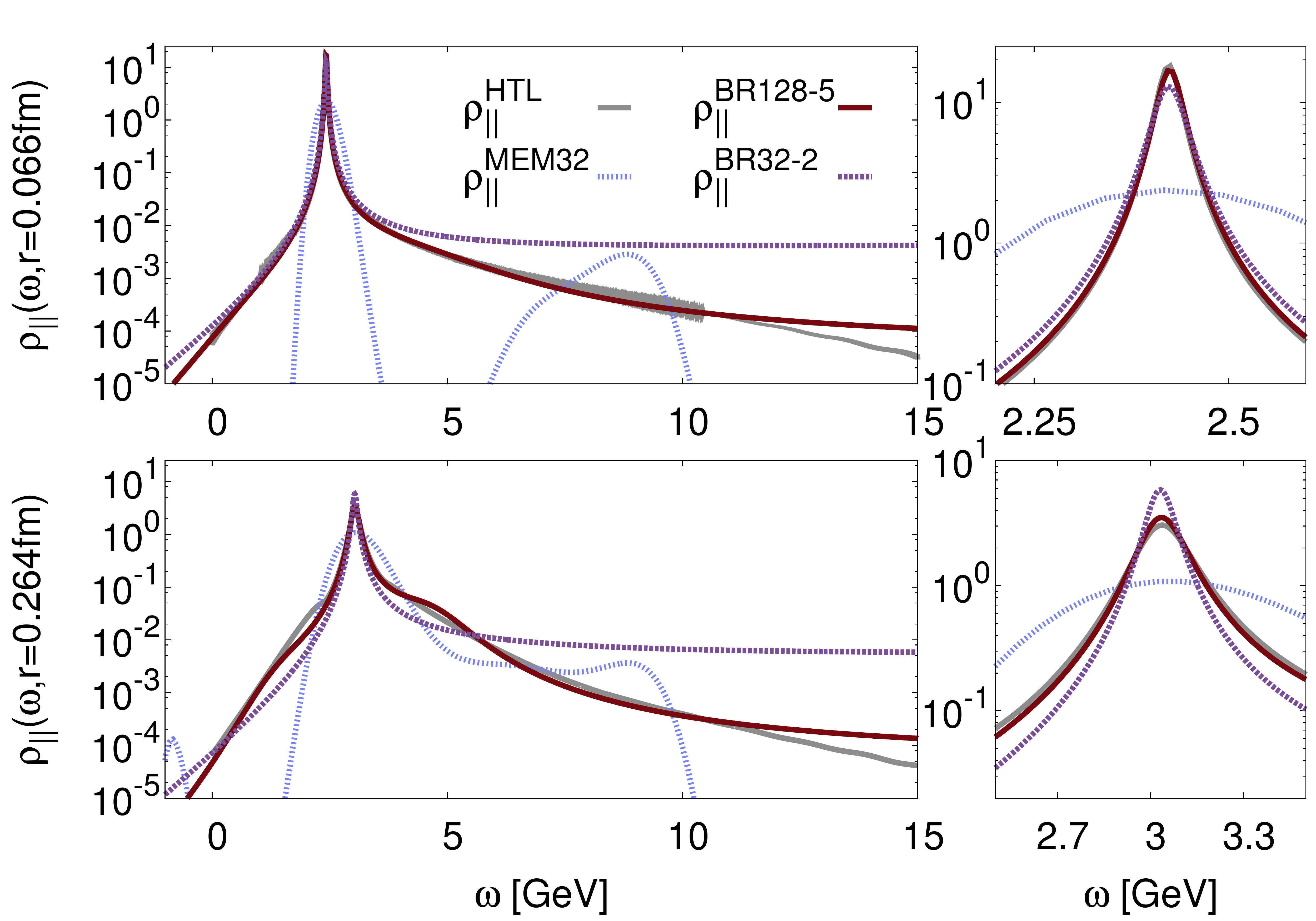}
 \caption{Comparison of the spectra: exact HTL, MEM with 32 points no errors added, our new method (BR) with 32 and 128 points and errors added \cite{Burnier:2013nla}. We show two distances, $r=0.06$fm and $r=0.26$fm (top): Wilson loop (respectively $10^{-4}, ~10^{-5}$ errors), (bottom): Wilson lines (respectively $10^{-2},~ 10^{-5}$ errors).}\label{comp}
\end{figure}

Independently of the method, the Wilson loop peak is harder to reconstruct because of the background from cusp divergences.
Comparing the methods, we see a clear advantage to the new bayesian reconstruction, which captures the main features of the spectrum very well, whereas the MEM only gets the main peak, however with a far too broad width and an incorrect shape. 

%%%%%%%%%%%%%%%%%%%%%%%
 \section{Potential}\label{s5}
 %%%%%%%%%%%%%%%%%%%%%%%
 
\begin{figure}[h]
\centering
\includegraphics[angle=-90,width=7.5cm]{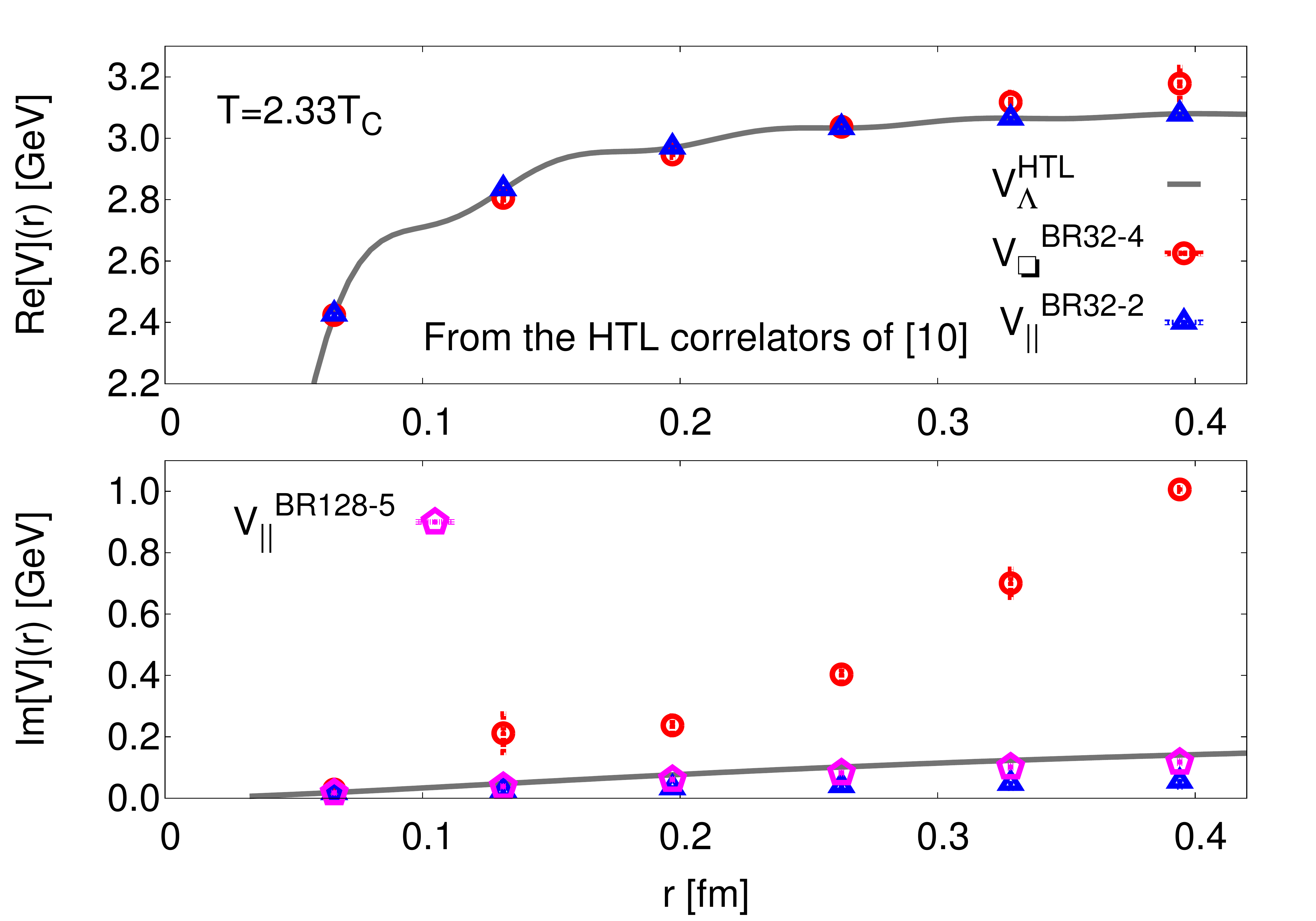}
\includegraphics[angle=-90,width=7.5cm]{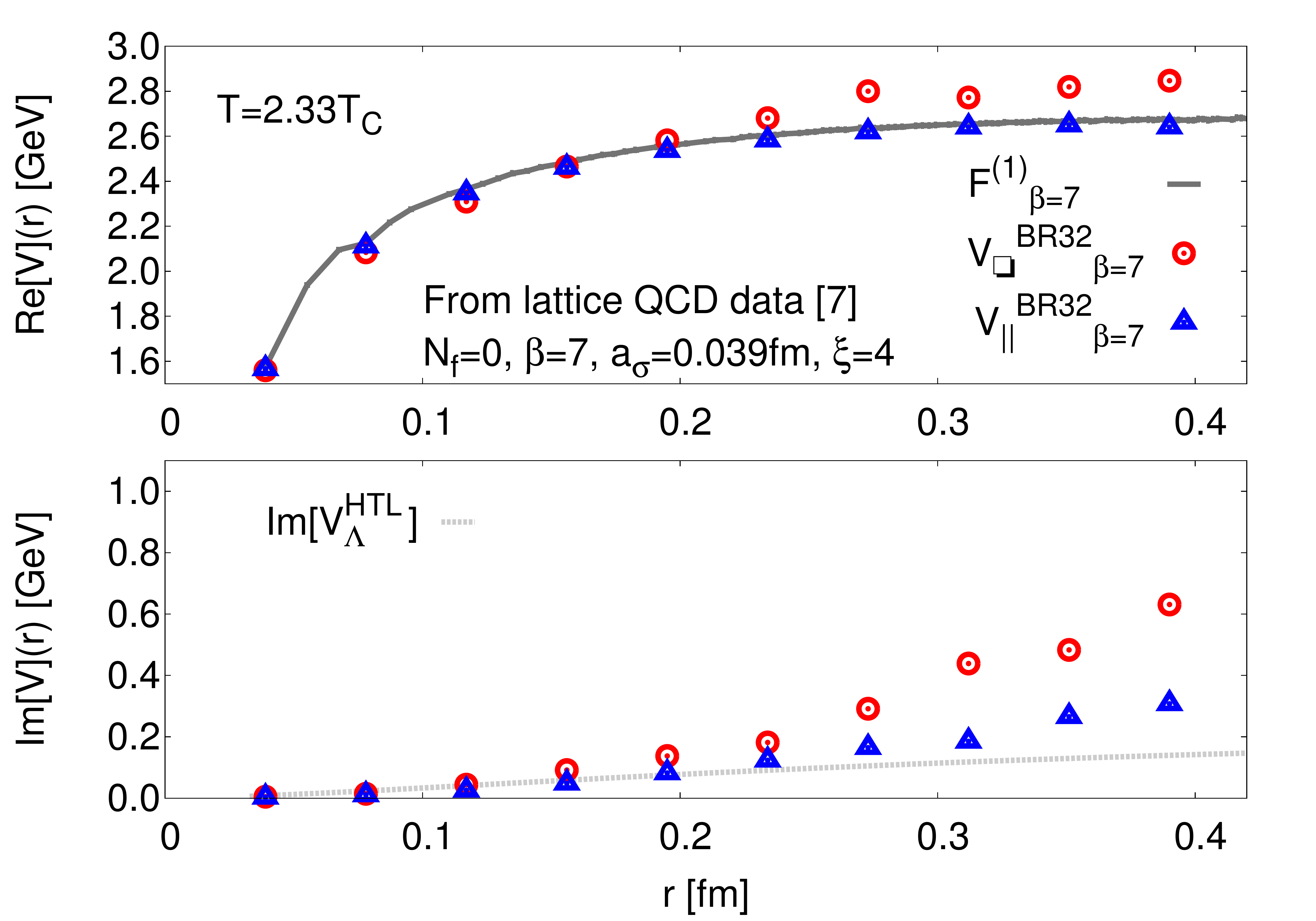}
\caption{Reconstruction of the HTL and quenched lattice QCD ($T=2.33T_C$) potential \cite{Burnier:2013nla}. (left) the HTL potential (solid line) based on the Euclidean HTL Wilson Loop $V_\square(r)$ (circle) and the HTL Wilson line correlator $V_{||}(r)$ (triangle). (top left): ${\rm Re}[V](r)$(bottom left): ${\rm Im}[V](R)$ requires $10^{-5}$ errors for a reliable determination. On the right: reconstruction of the quenched Lattice ($T=2.33T_C$, $N_\tau=32$,~$a=0.39$fm) potential based on the Euclidean Wilson Loop $V_\square(r)$ (circle) and the Wilson line correlator $V_{||}(r)$ (triangle). (top right): ${\rm Re}[V](r)$ together with the free energy (solid line). (bottom right): ${\rm Im}[V](R)$  with HTL potential.}\label{Figx} \vspace{-0.4cm}
\end{figure}

After testing the analytic continuation and the fitting procedure separately, we can now test the full procedure, doing first the numerical analytic continuation with our new method and then fitting the results with expression (\ref{Eq:FitShapeFull}). 
This is done as a check on the HTL data but also on the quenched lattice data of ref.~\cite{Rothkopf:2011db}. The resulting potentials are shown in fig.~\ref{Figx}.

We see that the real part of the potential obtained applying the full procedure on the HTL data matches the correct result very precisely for the Wilson lines (below one percent error already with data of $10^{-2}$ relative precision), whereas the extraction from the Wilson loop is not so precise at large $r$. For the imaginary part, a reliable extraction is only possible for very precise data and in the Wilson line case.
Turning to the real lattice data, we see that the unexpected rise of the potential obtained in ref.~\cite{Rothkopf:2011db} is fully gone, and is now understood as an artifact of MEM and of the naive fitting procedure used there. We see that the Wilson line data matches quite well, but not exactly, the free energy, as could be expected form the perturbative calculations of ref.~\cite{Burnier:2009bk}.

%%%%%%%%%%%%%%%%%%%
\section{Conclusion}\label{s6}
%%%%%%%%%%%%%%%%%%%%

The procedure to extract the potential from the lattice first proposed in ref.~\cite{Rothkopf:2011db} lead to counterintuitive results. The analysis of the lattice measurements yielded a potential that kept rising with $r$, even well above the phase transition temperature $T_c$. The HTL test allowed to clarify two issues:

If the lowest peak of the spectral function indeed encodes all information on the potential, it is not a Lorentzian as first assumed but a skewed Lorentzian. This result holds in a very general setup \cite{Burnier:2012az}, supposing that the $Q\bar Q$ system can be described by a potential at large times. Fitting a skewed peak by a Lorentzian overestimates the potential at large $r$, explaining in part the counterintuitive result of ref.~\cite{Rothkopf:2011db}.  Correcting this point lead to results compatible with Debye screening \cite{Burnier:2012az}.

The results based on extended MEM and the improved fit (\ref{Eq:FitShapeFull}) were however still inaccurate but a comparison of the reconstructed spectrum to the known HTL result shed light on the underling difficulties. The reconstructed width was far too broad and several additional peaks were present in the MEM reconstruction that where not encoded in the data. 
%The extended MEM thus enables a rough reconstruction of the real part of the potential but fails to capture the Lorentzian structure of the peak and the Imaginary part of the potential. It only gives hints to the existence of the large background in the Wilson loop case. 
To cure these issues we developed a new Bayesian approach that uses an unlimited search space to enable the reconstruction of very sharp peaks, that could in principle not be obtained with MEM simply because its functional basis is too restricted. We further changed the entropy functional so that the reconstruction is now forced to be smooth where the data do not constrain the spectrum, avoiding the generation of unphysical peaks.

With the new method, the real part of the potential can already be obtained with a precision below a precent by measuring the Wilson lines to better than a precent accuracy. This open up the use of the results for phenomenology. The imaginary part is however harder to reconstruct reliably and requires to measure the Wilson lines to $10^{-5}$ accuracy.

\section*{Acknowledgments}

This work was partly supported by the Swiss National Science Fundation (SNF) under grant 200021-140234 and under the Ambizione Grant PZ00P2-142524. The authors acknowledge the DFG-Heisenberg group of Y.~Schr\"oder at Bielefeld University for providing computer resources.

\end{document}